# Concept of the half-valley-metal and quantum anomalous valley Hall effect

He Hu[1], Wen-Yi Tong[1,†], Yu-Hao Shen[1], Xiangang Wan[2], Chun-Gang Duan[1,3,*]

[1]State Key Laboratory of Precision Spectroscopy and Key Laboratory of Polar Materials and Devices, Ministry of Education, Department of Electronics, East China Normal University, Shanghai, 200241, China

[2]Department of Physics and National Laboratory of Solid State Microstructures, Nanjing University, Nanjing 210093, China

[3]Collaborative Innovation Center of Extreme Optics, Shanxi University, Taiyuan, Shanxi 030006, China

**Abstract**

Valley, the energy extrema in the electronic band structure at momentum space, is regarded as a new degree of freedom of electrons, in addition to charge and spin. The studies focused on valley degree of freedom now form an emerging field of condensed matter physics, *i.e.* valleytronics, whose development is exactly following that of spintronics which focuses on the spin degree of freedom. Here, in analogy to half-metals in spintronics with one spin channel is conducting whereas the other is insulating, we propose the concept of half-valley-metal, in which conduction electrons are intrinsically 100% valley polarized, as well as 100% spin-polarized even when spin-orbit interactions are considered. Combining first-principles calculations with two-band **k·p** model, the physical mechanism to form the half-valley-metal is illuminated. Taking the ferrovalley H-$FeCl_2$ monolayer with strong exchange interaction as an example, we find that the strong electron correlation effect can induce the ferrovalley to half-valley-metal transition. Due to the valley-dependent optical selection rules, such system could be transparent to, e.g., left-circularly polarized light, yet the right-circularly polarized light will be reflected, which can in turn be used as a crucial method to detect half-valley-metal state. In addition, we find that in the so obtained half-valley-metal state, the conduction valley demonstrates Dirac cone-like linear energy dispersion. Interestingly, with the increase of the correlation effect, the system becomes insulating again with all valleys follow same optical selection rule. We confirm that in this specific case, the valence bands, which consist of single spin, possess non-zero Chern number and consequently *intrinsic* quantum anomalous valley Hall effect emerges. Our findings open an appealing route toward functional 2D materials design of valleytronics.

**Introduction**

In recent years, since the emergence of novel two-dimensional (2D) graphene-related materials [1-4] with hexagonal lattice symmetry, a new degree of freedom of electrons, i.e. valley [5, 6], has attracted intensive attention. In analogy to charge and spin, the valley degree of freedom constitutes the binary logic states in solids, leading to possible applications for information processing. The field of valleytronics has flourished with the study of H-phase group-VI transition metal dichalcogenides (TMDs) [7-10], due to their inequivalent $K_+$ and $K_-$ valleys [11]. Till now, various functional devices have been achieved in valleytronics, such as valley separator, valley filter, valley valve, electron beam splitter and logic gate [12-16]. Yet, the energy degeneracy of the two prominent valleys does not meet the requirement of nonvolatility in next-generation information technology. Efforts in valley degeneracy lifting have been made in TMDs using external fields [17-30]. Recently, the concept of ferrovalley materials, as a new member of the ferroic family, was proposed. The existence of spontaneous valley polarization, resulting from either ferromagnetism in hexagonal 2D materials [31] or ferroelectricity with orthorhombic lattice [32] renders nonvolatile valleytronics applications realizable.

The development of valleytronics is indeed very similar to that of spintronics. As is well known, in the field of spintronics, half-metals [33], where one spin channel is metallic while the other is semiconducting or insulating, can provide completely spin-polarized currents by filtering the current into a single spin channel without any external operations. Therefore, such states are of great importance in both theoretical study and practical applications of spintronics. For valleytronics, the analogous field of spintronics, the following questions are naturally raised: whether there is a half-valley-metal state, in which electrons are of metallic characters at one valley whereas at the opposite valley they keep semiconducting? If the answer is yes, is there any novel property associated with this half-valley-metal?

Furthermore, the quantum valley Hall effect, rooted in the spatial noncentrosymmetry, has been extensively studied in various valley systems [34-37]. Through additionally introducing the time-reversal symmetry broken, the valley-polarized quantum anomalous Hall effect has been proposed in model calculations [38] and then demonstrated in first-principles calculations [39], which combines valleytronics and topology, two hot topics in condensed-matter physics, together. However, valley Hall current is absent in the proposed systems, making the detection of the Hall effect difficult. Another question then arises, is it possible to find a natural system with both the quantum anomalous Hall effect and the valley index, or in another term, quantum anomalous valley Hall effect (QAVHE)?

In this Letter, starting from the two-band **k·p** model, we successfully gain the desired half-valley-metal state. Then we achieve this new state in ferrovalley H-FeCl$_2$ monolayer, which possesses strong spin-orbit coupling (SOC) effect and large exchange interaction of transition-metal-*d* electrons, by tuning the on-site Coulomb interaction. The electrons around the Fermi level of the system are 100% valley and 100% spin polarized with Dirac-cone-like linear dispersion. More interestingly, the QAVHE is revealed in H-FeCl$_2$ monolayer with appropriate on-site Coulomb interaction, while the valley-related special optical properties are still remained.

***k · p* analysis**

Similar to H-VSe$_2$ with hexagonal structure [31], a two-band **k·p** model including magnetic interaction is employed here to describe the electronic properties near the K$_\pm$ in the 2D Brillouin zone (BZ) of monolayers of magnetic H-phase TMDs. The conduction state $\psi_1^\tau = |d_{z^2}\rangle$ and the valence state $\psi_2^\tau = (|d_{x^2-y^2}\rangle + i\tau|d_{xy}\rangle)/\sqrt{2}$ are chosen as the basis functions with index $\tau = +1$ (-1) denotes the point K$_+$ (K$_-$). Then, the total effective Hamiltonian of the system is given as:

$$H(\mathbf{k}) = \begin{bmatrix} \frac{\Delta}{2} + \varepsilon + \sigma\tau\lambda_c - sm_c & t(\tau q_x - iq_y) \\ t(\tau q_x + iq_y) & -\frac{\Delta}{2} + \varepsilon + \sigma\tau\lambda_v - sm_v \end{bmatrix}, \quad (1)$$

here $\Delta$ is the band gap of the TMD monolayer without magnetism and the SOC effect at the valleys K$_\pm$, $\varepsilon$ is a correction energy bound up with the Fermi level, $t$ is the effective nearest-neighbor hopping integral, and $q = k - K_\pm$ is the relative momentum vector with respect to the valleys K$_\pm$. $\lambda_{c(v)}$ and $m_{c(v)}$ respectively demonstrate the spin splitting derived from the SOC effect and the effective exchange splitting at the bottom of conduction band (CB) (the top of valence band, VB). Spin is indexed by $\sigma$ with +1 (–1) for spin-up (down) state.

When we focus on the spin down channel, the spin index in the SOC and exchange interaction terms is -1. By diagonalizing the above Hamiltonian, we obtain the energy spectra:

$$E(\psi^\tau) = \frac{1}{2}\left[ 2\varepsilon - \tau(\lambda_c + \lambda_v) + (m_c + m_v) \pm \sqrt{(\Delta')^2 + 4t^2(q_x^2 + q_y^2)} \right] \quad (2)$$

Here we introduce the gap parameter $\Delta' = \Delta - \tau(\lambda_v - \lambda_c) + m_c - m_v$, which corresponds to the energy difference between $|\psi_1^\tau\rangle$ and $|\psi_2^\tau\rangle$ states at the two valleys ($q_x = q_y = 0$). For a single valley, through tuning the SOC ($\tau\lambda_{v(c)}$) and the exchange effect ($m_{c(v)}$), $\Delta'$ could be positive, zero and even negative. Note here a negative $\Delta'$ indicates an interesting state with inverted band ordering, i.e. $E(\psi_1^\tau) < E(\psi_2^\tau)$. Whereas for $\Delta' = 0$, the metallic state without energy gap could be expected. In addition, it is interesting to point out that the valley splitting, i.e., energy difference between the band gap at K$_+$ ($E_g^{K_+} = \Delta + \lambda_v - \lambda_c + m_c - m_v$) and K$_-$ ($E_g^{K_-} = \Delta - \lambda_v + \lambda_c + m_c - m_v$), possesses the magnitude of $2|\lambda_v - \lambda_c|$, merely depending on the SOC effect. As an element-sensitive effect, the SOC term is generally hard to be affected. The valley splitting is thus treated as a constant in the following discussions.

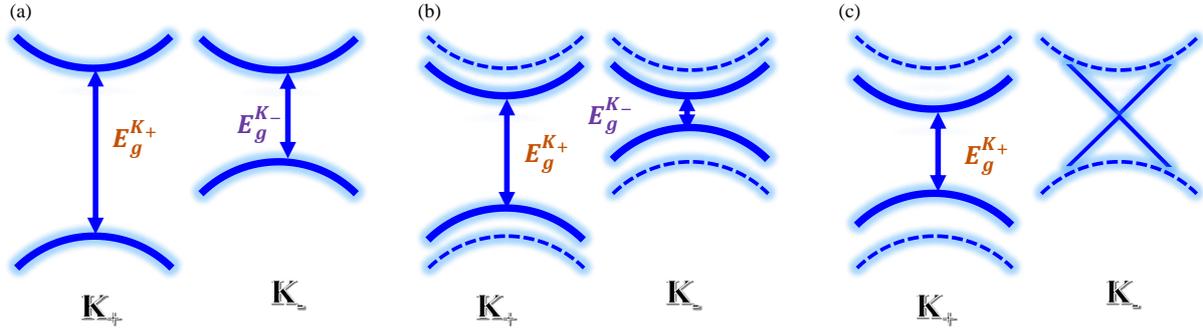

Figure 1. The schematic band structures near valleys $K_+$ and $K_-$ for ferrovalley states with (a) larger and (b) smaller gap, (c) half-valley-metal states with one of the valley has zero gap. $E_g^{K_+}$ ($E_g^{K_-}$) represents the allowed interband transitions excited by circularly polarized light at $K_+$ ($K_-$) valley with only spin down channel.

Starting from the schematic band structures of traditional ferrovalley materials, as shown in Fig. 1(a), we find that the band gap of the two valleys can be regulated by the exchange interaction ($|m_c - m_v|$) synchronously. When we decrease the band gap at $K_+$ valley, the one at $K_-$ will be reduced as well (shown in Fig. 1(b)). Such kind of simultaneous band gap changing provides the possibilities to obtain the desirable critical state, as shown in Fig. 1(c), of which at one valley the gap is closed while at the other one the gap is still open. Interestingly, for $K_-$ valley with $\Delta'=0$, it owns a linear dispersion $E(\psi^\tau)=\frac{1}{2}[2\varepsilon-\tau(\lambda_c+\lambda_v)+(m_c+m_v)\pm 2tq]$, similar to that of the Dirac cone. Yet, it corresponds to a single spin channel.

Here, we name the state like Fig. 1(c) as half-valley-metal, where one valley presents metal properties and the other is still in semiconductor status. Note that due to the unbroken symmetry, the optical selection rule [36] is still held, i.e. the circularly-polarized light is still locked with valley chirality. To be specific, for $K_+$ valley with gap, only left-circularly-polarized light can be absorbed. An interesting phenomenon thus emerges: when we use the left-circularly-polarized light to make detection, we find the system has finite bandgap. Whereas for the right-circularly-polarized one, the systems will reflect the light just like metal does. We would like to mention that besides the one based on the optical selection rule, other peculiar properties, such as 100% spin-polarized conduction electron, are expected for such half-valley-metal state, since it is born of the half-metallic one. In addition, since the SOC effect is already considered in the above analysis, the 100% spin polarization is robust. This is in strong contrast to the traditional half metal [33], in which the SOC effect will generally mix the spin-up and spin-down states, making the system not exactly 100% spin polarized. Whereas for the normal Dirac half metal [40], characterized by a band structure with a gap in one channel but a Dirac cone in the other, the SOC effect always opens a band gap and eventually makes the system going to be semiconductor. Therefore, the half-valley-metal may be a true Dirac half metal with 100% spin polarization.

In order to explore the Hall effect, we then calculate the Berry curvature of the system. Here, we consider the spin-resolved Berry curvature from the Kubo formula [41]:

$$\Omega_n^{\uparrow(\downarrow)}(\mathbf{k}) = -\sum_{n' \neq n} \frac{2 \operatorname{Im} \langle \varphi_{n,\mathbf{k}}^{\uparrow(\downarrow)} | v_x | \varphi_{n',\mathbf{k}}^{\uparrow(\downarrow)} \rangle \langle \varphi_{n',\mathbf{k}}^{\uparrow(\downarrow)} | v_y | \varphi_{n,\mathbf{k}}^{\uparrow(\downarrow)} \rangle}{(E_{n'}^{\uparrow(\downarrow)} - E_n^{\uparrow(\downarrow)})^2}, \tag{3}$$

The summation is over all the occupied states. Here, $E_n$ represents the eigenvalue of the Bloch function $\varphi_{n,\mathbf{k}}^{\uparrow(\downarrow)}$, while both $v_x$ and $v_y$ are velocity operators in the $x$ and $y$ directions, defined as $v_x = \frac{\partial H}{\partial k_x}$ and $v_y = \frac{\partial H}{\partial k_y}$. In terms of the **k·p** model Hamiltonian parameterization Berry curvature can be simplified as:

$$\Omega_z(\boldsymbol{k}) = \frac{2t^2 \tau \Delta'}{[4t^2(q_x^2 + q_y^2) + (\Delta')^2]^{\frac{3}{2}}}, \tag{4}$$

Integrating the Berry curvature over the first BZ in the 2D system gives the total Chern number $C = \frac{1}{2\pi} \sum_n \int_{BZ} d^2\boldsymbol{k} \Omega_z(\boldsymbol{k})$ [42].

From Eq. (4) we can easily know that the Berry curvature is significant only around the valley and its sign is determined by the sign of the product of $\tau$ and $\Delta'$. Note that in ordinary valley systems, $\Delta'$ is generally positive, though may be different as in the case of ferrovalley state, at K$_+$ and K$_-$ valleys. This indicates that the Berry curvatures have opposite signs at the two valleys, as they possess different $\tau$. Consequently, the Chern number is generally zero due to the opposite contribution to the integration of $\Omega_z$ from the two valleys. This immediately brings a conjecture. As we mentioned earlier, $\Delta'$ could even be negative with appropriate $m_c$ and $m_v$. If we further increase $|m_c - m_v|$, the system will stride over the half-valley-metal state, and $\Delta'$ at K$_-$ will change from zero to negative value while at K$_+$ it is still with positive sign. At this moment, the product $\tau \Delta'$ has the same positive sign at K$_+$ and K$_-$ valleys. By integrating the Berry curvature in the first BZ, we should get a *nonzero* Chern number, suggesting now the system is in the topological state. And, as a result, we expect the QAVHE will occur in this system.

*Ab initio* **calculations**

In order to verify the results anticipated by the **k·p** model, we then carry out *ab initio* calculations. We choose H-FeCl$_2$ as our sample system, as Fe ions generally demonstrate larger exchange splitting. Fig. 2(a) and Fig. 2(b) show its geometric structures. In such a monolayer, an intermediate layer of hexagonally arranged Fe atoms is sandwiched between two layers of Cl atoms. Each Fe atom is surrounded by six Cl atoms. All calculations are performed in the framework of density functional theory using the Vienna Ab Initio Simulation Package (VASP) [43]. Interactions between valence electrons and ionic cores are described with the projector augmented wave method [44]. The generalized gradient approximation with Perdew–Burke–Ernzerh parametrization was applied [45]. A plane wave basis set with a cutoff energy of 600 eV is used to expand the wave functions. The 2D BZ integration is sampled by an 18 × 18 × 1 $k$-grid mesh for calculations of electronic properties. A vacuum layer with a thickness of 20 Å is set in the calculation for monolayer FeCl$_2$ to ensure decoupling between periodic FeCl$_2$ layers. For structural relaxation, all the atoms are allowed to relax until the atomic force on each atom is smaller than 0.001 eV Å$^{-1}$. Both the lattice constant and the atomic positions is fully optimized. The SOC effect [46] is explicitly included in the calculations.

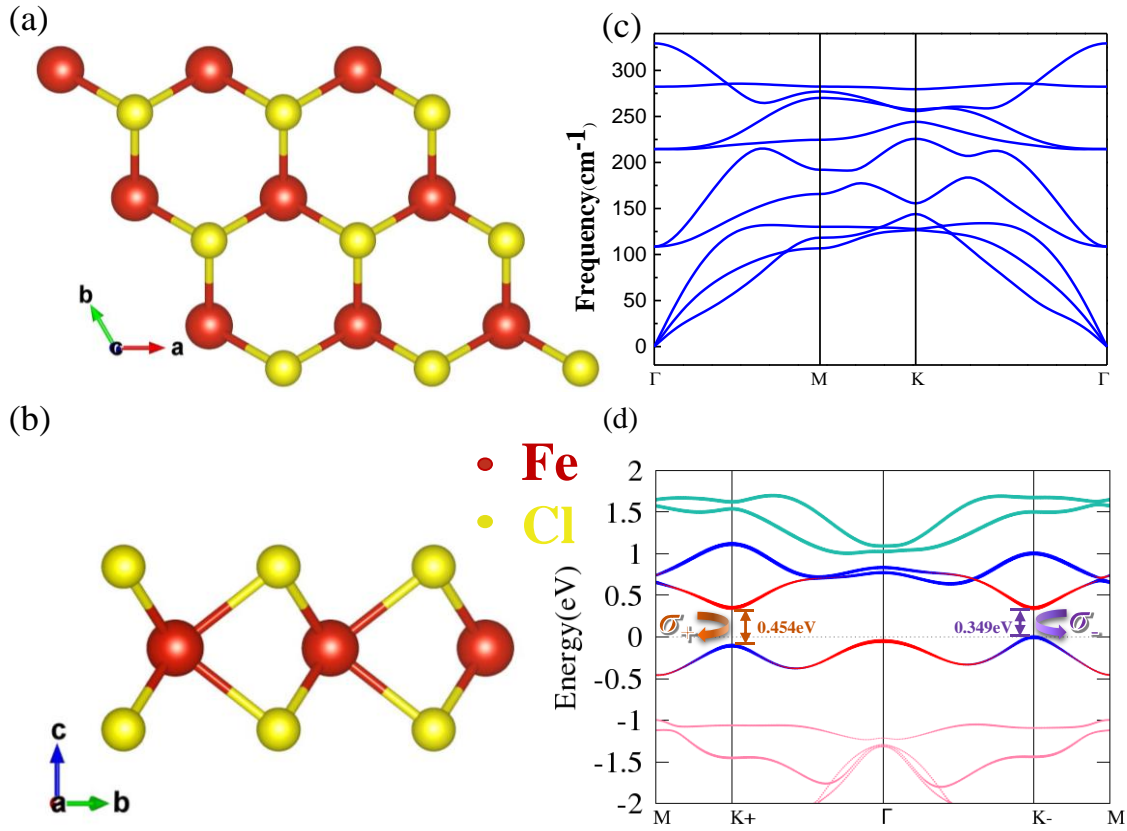

Figure 2. (a) Top and (b) side view of H-FeCl$_2$ monolayer. (c) Phonon spectrum of H-FeCl$_2$ monolayers. (d) Orbital projected band structures of H-FeCl$_2$ monolayer with $U_{eff}$ =0 eV. The radius of dots is proportional to its population in corresponding states: blue symbols represent spin-down components of both Fe-$d_{xy}$ and Fe-$d_{x^2-y^2}$ orbitals on cation-Fe, red ones for spin-down components of Fe-$d_{z^2}$ orbitals, green ones for both Fe-$d_{xz}$ and Fe-$d_{yz}$ characters, and pink symbols represent spin-up Fe-$d$ states. σ+ and σ- represent the left-handed and right-handed radiation, respectively.

To assess the dynamical stability of the H-FeCl$_2$ monolayer under different strains, we calculate their phonon dispersion by using the Quantum-ESPRESSO [47] code. For the optical property calculation, we adopt our own code OPTICPACK, which has been successfully used in various systems [48, 49]. To describe the strong-correlated effect of Fe-3$d$ electrons, the Dudarev's approach of the LSDA + $U$ scheme is adopted [50], in which only the effective $U$ ($U_{eff}$) based on the difference between the on-site Coulomb interaction parameter ($U$) and exchange parameters ($J$) is meaningful. We further calculated the Chern number and edge states with the software package Wannier tools [51] using the renormalized effective tight binding Hamiltonian.

The optimized lattice constant is 3.364 Å, and the angle of Cl–Fe–Cl is 76.6 degrees. The magnetic moment of Fe is calculated to be 4.00 $\mu$B, in consistent with previous theoretical results of H-FeCl$_2$ [52] and T-FeCl$_2$ [53]. In its bulk phase, FeCl$_2$ is reported to be an antiferromagnetic insulator with space group $P\bar{3}m1$, consisting of a sequence of -Cl-Fe-Cl-Cl-Fe-Cl- layers where the Cl-Cl layers are bound by weak Van der

Waals forces [54]. At present, there is no related experimental report about the monolayer FeCl$_2$. To determine the dynamical stability of H-FeCl$_2$ monolayer, its phonon spectrum along the high symmetry lines within the first BZ is shown in Fig. 2(c). The absence of imaginary frequencies demonstrates the stability of FeCl$_2$ monolayer in H phase, which is in agreement with other theoretical work [52].

The valley-shaped dispersion near K$_+$ and K$_-$ points is clearly seen in the band structures of the FeCl$_2$ monolayer (Fig. 2(d)). The spin-down Fe-3$d_{z^2}$ orbitals occupy the bottom of CB, while spin-down Fe-3$d_{xy}$ and Fe-3$d_{x^2-y^2}$ characters hold the top of VB at the two k points. Spin-down components of Fe-$d_{xz}$ and Fe-$d_{yz}$ locate at higher energy range. For the spin-up Fe-$d$ states, they are in the energy range below -1 eV, far away from the Fermi level. Furthermore, we analyze the symmetries of K$_+$ and K$_-$ points. The irreducible representations (IRs) at K$_+$ are 2E$_{1/2}$ and 2E$_{5/2}$ for the top VB and bottom CB, respectively. At K$_-$, however, they are 2E$_{1/2}$ (top VB) and 2E$_{3/2}$ (bottom CB). The different symmetry for bottom CB makes the K$_+$ and K$_-$ points unequal. When the left handed incident light (2E′ symmetry) with frequency ~ ($E_g/\hbar$) comes, according to the great orthogonality theorem that the electric-dipole transition is allowed only if the reduced direct product representation between the initial state IRs and the incident light IRs contains the representation of the final state, we can obtain that: 2E$_{1/2}$ ⊗ 2E′ = 2E$_{5/2}$. Therefore it can only be absorbed at K$_+$ point. Similarly, the optical absorption at the K$_-$ point can be excited merely by the right-handed circularly polarized light (1E′ symmetry) due to the relation 2E$_{1/2}$ ⊗ 1E′ = 2E$_{3/2}$. Such dependence of optical selection rules on k points signifies that K$_+$ and K$_-$ points in H-FeCl$_2$ monolayer can be regarded as two prominent valleys.

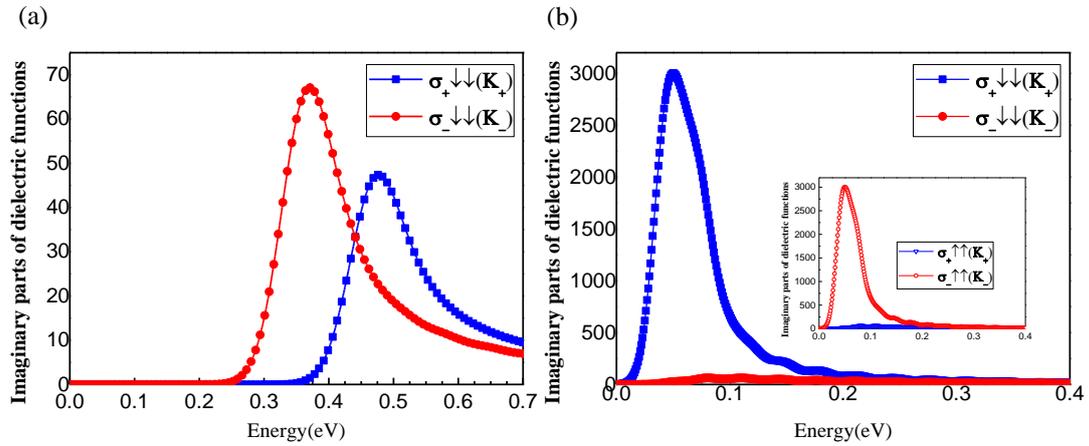

Figure 3. The imaginary parts of complex dielectric function $\varepsilon_2$ for monolayer H-FeCl$_2$. (a) $U_{eff}$ = 0 eV (with primary valley polarization) and (b) $U_{eff}$ = 1.80 eV. Red (blue) symbols with lines represent right-handed (left-handed) polarized light. The insert in (b) corresponds to the system with negative magnetic moment.

Interestingly, thanks to the coexistence of ferromagnetism and SOC effect, two valleys here are intrinsically nonequivalent. The valley splitting, defined as $\Delta E = |E_g^{K_+} - E_g^{K_-}|$, where $E_g^{K_\pm}$ is the band gap at K$_\pm$ valley, becomes 106 meV, which is a little bit larger than that in VSe$_2$ (90 meV) [31]. Such valley splitting is also exactly represented in the 106 meV red shift of the optical band gap of right-handed radiation, compared with that of the left-handed one, as is shown in Fig. 3(a). The chirality-dependent optical band gap proves H-FeCl$_2$ monolayer as an intrinsic ferrovalley material.

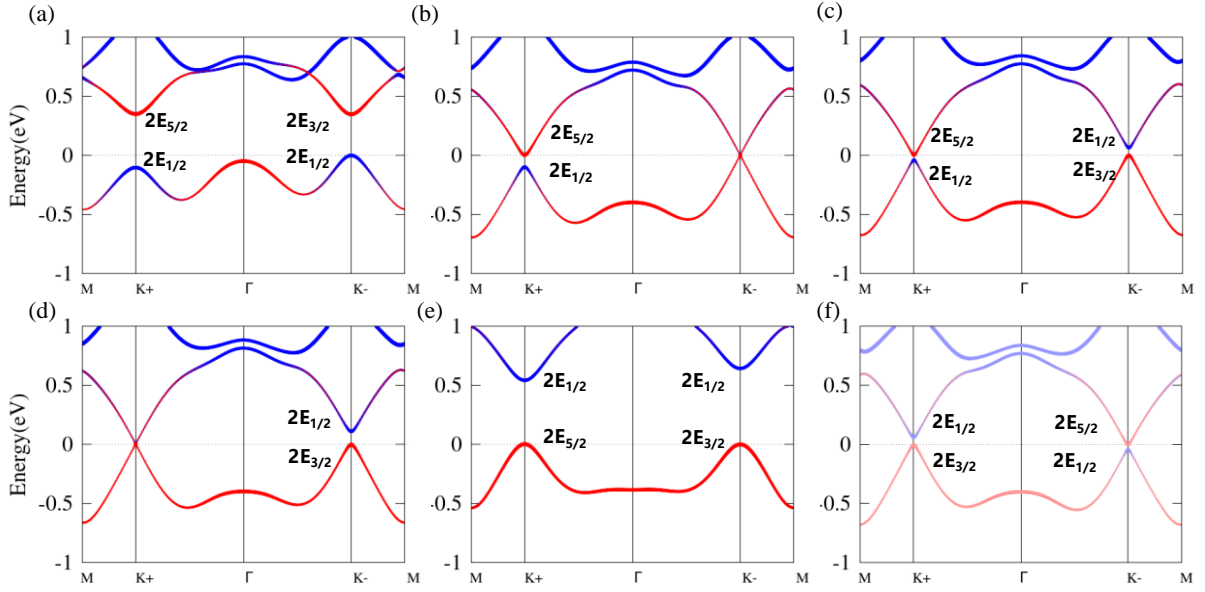

Figure 4. Band structures of H-FeCl$_2$ monolayer. (a) Ferrovalley state with $U_{eff}$ = 0 eV, (b) half-valley-metal state with $U_{eff}$ = 1.67 eV, (c) topologically nontrivial state with $U_{eff}$ = 1.80 eV, (d) half-valley-metal state with $U_{eff}$ = 1.90 eV, (e) Ferrovalley state with $U_{eff}$ = 3 eV, (f) the same as (c) but with opposite magnetic moment. Blue symbols represent spin-down components of Fe-$d_{xy}$ and Fe-$d_{x^2-y^2}$ orbitals. Red ones are spin-down states for Fe-$d_{z^2}$ characters. Light blue (light red) is for spin-up components of Fe-$d_{xy}$ and Fe-$d_{x^2-y^2}$ orbitals. The IRs of states have been labelled using the Mulliken notations.

Previous analysis starting from the minimal **k·p** model proposes the feasibility of achieving the half-valley-metal states and the QAVHE. For the H-FeCl$_2$ monolayer, the effective $U$ parameter ($U_{eff}$) is adopted to describe the strong-correlated effect of electrons in 3$d$ shell of Fe atoms. Fig. 4(a), as a reference, presents the original state of H-FeCl$_2$ without $U$. When an $U_{eff}$ is applied, VB occupied by Fe-$d_{xy}$ and Fe-$d_{x^2-y^2}$ moves up, Fe-$d_{z^2}$ dominated CB correspondingly goes down. Once $U_{eff}$ reaches up to 1.67 eV, as shown in Fig. 4(b) the band gap gets closed initiatively at K$_-$ valley, meanwhile a narrow band gap is kept at K$_+$ valley, signifying the half-valley-metal status. Note that the band structure at the K$_-$ valley now demonstrates the Dirac cone shaped linear dispersion, perfectly reproduced the characteristic anticipated by our **k·p** model. It provides mass-free electron mobility, in favor of charge and spin transport. After the critical state we further increase $U_{eff}$. The Fe-$d_{x^2-y^2}$ and Fe-$d_{xy}$ occupied bands continue to move up, while bands of the Fe-$d_{z^2}$ components go down at both of the two valleys. For the case in Fig. 4(c) with $U_{eff}$ = 1.80 eV, the CB and VB are still in the process of getting closed at K$_+$ valley. Yet at the K$_-$ valley, the CB and VB are apart from each other. Here the components of bottom CB and top VB at K$_-$ are completely exchanged compared with that of the original H-FeCl$_2$ monolayer state. Now, the IRs at K$_-$ change to 2E$_{3/2}$ and 2E$_{1/2}$ for the top VB and bottom CB, respectively. The transition relationship 2E$_{3/2}$ ⊗ 2E′= 2E$_{1/2}$ corresponds to the left-handed light absorption at K$_-$ valley. The same symmetry at K$_+$ and K$_-$ valleys indicating that only left-handed light can be absorbed in this state. When the effective $U$ is up to 1.90 eV, we can gain the other half-valley-metal state (Fig. 4(d)), where the K$_-$ valley holds the optical visible band gap and the K$_+$ valley is metallic, opposite to the case shown in Fig. 4(b). Further increasing $U_{eff}$ makes the components of CB and VB at K$_+$ valley reversed as well. Eventually, the system turns back to the ferrovalley state again. Fe-$d_{xy}$ and Fe-$d_{x^2-y^2}$ occupied VB start to move up while Fe-$d_{z^2}$ occupied CB goes down which means the system starts to turn back to the ferrovalley state. We choose $U_{eff}$ =3 eV as an example (Fig. 4(e)), the components Fe-$d_{xy}$ and Fe-$d_{x^2-y^2}$ occupy conduction bands, while Fe-$d_{z^2}$ starts take over valence bands. Though the $K_+$ valley absorbs right-handed polarized light while $K_-$ takes left-handed polarized light, the narrow optical band gap absorption still corresponds to right-handed polarized light, indicating the redistribution of the two valleys compared with that of without $U_{eff}$.

Between the two half-valley-metal states, both two valleys correspond to the left-handed polarized light absorption. We calculate the circular polarized light absorption with applied 1.80 eV $U$ values in Fig. 3(b). It is clear to see that the absorption peak completely comes from the left-handed light, which is of great benefit to design the optical filter. The inset map corresponds to the inverted magnetic moment system, of which band structure clearly displayed in Fig. 4(f). As expected, the system will absorb right-handed light when the magnetization and then valley polarization reverses, which is confirmed by our calculation. This unique property suggests that, when the normal light incidents into the material, only the specific circularly polarized light can be absorbed and its fluorescence spectrum is also chirality dependent.

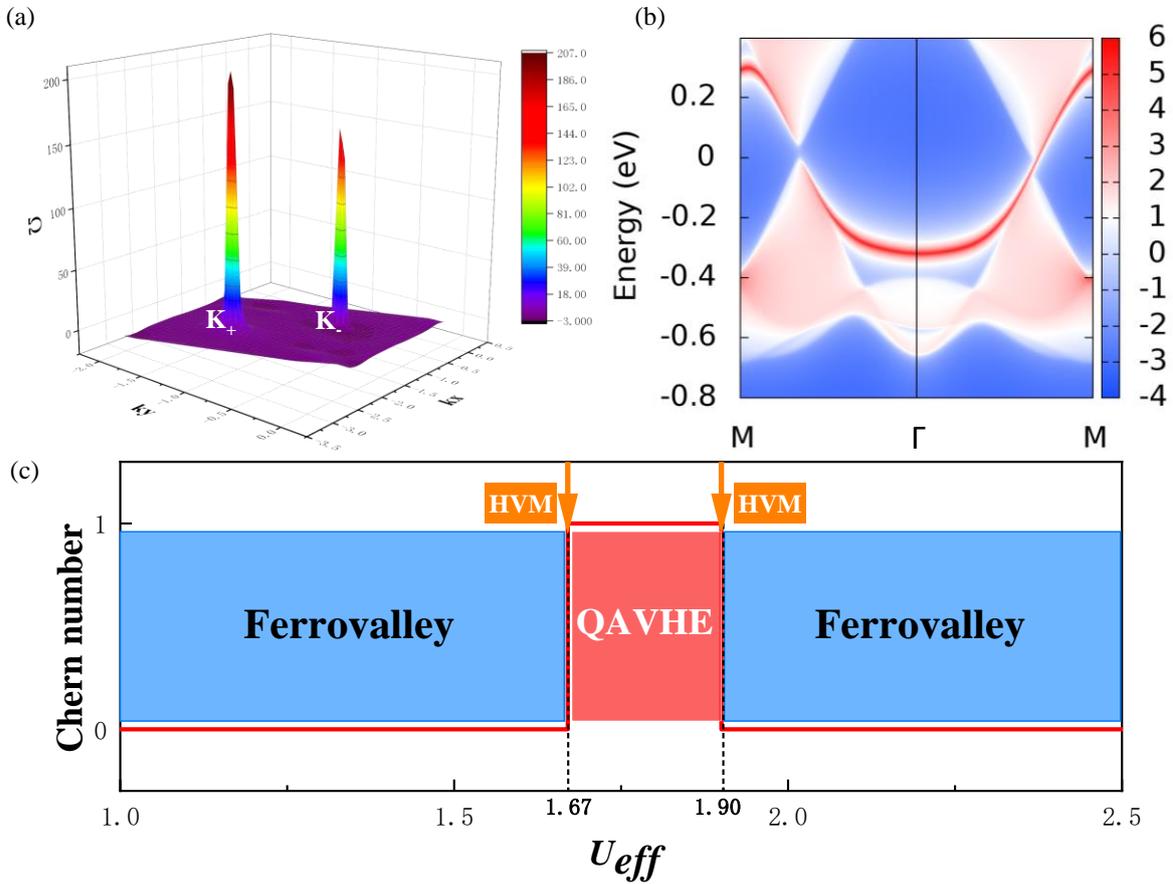

Figure 5. (a) Berry curvature in the momentum space in arbitrary units with $U_{eff}$ =1.80 eV. (b) Edge state along the zigzag direction. (c) Phase diagram of monolayer FeCl$_2$ with varied $U_{eff}$. The half-valley metallic state (shown as HVM in the figure as the down arrow indicated) is between the ferrovalley state and the topological nontrivial state which can demonstrate quantum anomalous valley Hall effect (QAVHE).

The Berry curvature and Chern number of the monolayer FeCl$_2$ is calculated with varied $U$ value from 0 eV to 2.5 eV and the topological phase diagram is shown at Fig. 5(c). Taking $U_{eff}$ =1.80 eV as an example, we obtain the Berry curvatures for the whole VBs of the first BZ in Fig. 5(a). The calculated Berry curvatures are almost zero except those points around the $K_+$ and $K_-$ valleys, as expected. In addition, they are of the same sign near two valleys, just the values are different. Integration over the Berry curvatures gives a nonzero Chern number (=1). This confirms our previous conjecture and indicates the existence of topological edge states within the insulating bulk gap.

We then calculated the edge states of a semi-infinite system of $FeCl_2$ with a zigzag edge. As shown in Fig. 5(b), there does exist the edge state connecting the conduction and valence bands, which is consistent with the Chern number calculation. Conclusively, among the two half-valley-metal states the nontrivial topological state is relative robust. The QAVHE state is kept while effective $U$ values varying from 1.67 eV to 1.90 eV. In our case, the system with the QAVHE retains the valley degree of freedom, therefore brings quantum anomalous Hall effect into valleytronics.

**Conclusion**

In summary, combining first-principles calculations with two-band **k·p** model we demonstrate the possibility of realizing the new ferrovalley member, half-valley-metal, which allow the electrons conduct at one valley whereas display the semiconducting states at the other valley. This is indeed a critical state of the valley system, where the conduction electrons are fully spin and valley polarized. It demonstrates dramatic chirality-dependent optical properties. Interestingly, the system can evolve from topologically trivial to nontrivial state with nonzero Chern number. In this topological valley state, the QAVHE with valley degree of freedom is confirmed, which ensures the non-dissipative high-performance electrons transportation in valleytronics. We hope these new concepts will enrich our understanding about valleytronics, which may accelerate its applications in the field of information processing and optoelectronics.

**Acknowledgements**

This work was supported by the National Key Research and Development Program of China (2017YFA0303403), Shanghai Science and Technology Innovation Action Plan (No. 19JC1416700), the NSF of China (No. 51572085, 11774092), ECNU Multifunctional Platform for Innovation.

[†:] twy891137ssyy@163.com

[*:] cgduan@clpm.ecnu.edu.cn